# Manipulating *dc* currents with bilayer bulk natural materials

Tiancheng Han[1#], Huapeng Ye[1#], Yu Luo[2#], Swee Ping Yeo[1], Jinghua Teng[3], Shuang Zhang[4], and Cheng-Wei Qiu[1]*

[1] *Department of Electrical and Computer Engineering, National University of Singapore, 119620, Singapore*

[2] *Department of Physics, Blackett Laboratory, Imperial College London, London SW7 2AZ, UK*

[3] *Institute of Materials Research and Engineering, Agency for Science, Technology and Research, 3 Research Link, Singapore 117602, Singapore*

[4] *School of Physics and Astronomy, University of Birmingham, Birmingham B15 2TT, UK*

*Corresponding author: chengwei.qiu@nus.edu.sg    # Equal contribution

**Keywords:** Cloaking sensor, *dc* bilayer cloak, fan-shaped concentrator, *dc*-current manipulation, homogeneous and isotropic materials

**Abstract:** The principle of transformation optics has been applied to various wave phenomena (e.g., optics, electromagnetics, acoustics and thermodynamics). Recently, metamaterial devices manipulating *dc* currents have received increasing attention which usually adopted the analogue of transformation optics using complicated resistor networks to mimic the inhomogeneous and anisotropic conductivities. We propose a distinct and general principle of manipulating *dc* currents by directly solving electric conduction equations, which only needs to utilize two layers of bulk natural materials. We experimentally demonstrate *dc* bilayer cloak and fan-shaped concentrator, derived from the generalized account for cloaking sensor. The proposed schemes have been validated as exact devices and this opens a facile way towards complete spatial control of *dc* currents. The proposed schemes may have vast potentials in various applications not only in *dc*, but also in other fields of manipulating magnetic field, thermal heat, elastic mechanics, and matter waves.

Controlling electromagnetic (EM) fields so as to render an object invisible, has been a long-standing dream for many researchers over the decades [1,2]. On the basis of the invariance of Maxwell's equations where equivalence is established between metric transformations and changes in material parameters, transformation optics [3] and conformal mapping [4] have been developed to manipulate EM wave propagation in a practically arbitrary manner. Besides making objects invisible [1-4], many other novel devices are rapidly emerging, with a representative one being a concentrator [5,6] that can enhance the energy density of incident waves in a given area. In addition to manipulation of EM waves [1-6], the theoretical tool of coordinate transformation has been extended to other areas of physics (such as acoustic waves [7], matter waves [8] and elastic waves [9]).

Recently, many significant achievements have been made in the manipulation of magnetostatic field [10-15], thermal conduction [16-19], and electrostatic field [20-24]. In 2007, Wood and Pendry proposed a *dc* metamaterial that pointed the way towards the design of static magnetic cloak [10], and the *dc* metamaterial was experimentally verified soon afterwards [11]. Recently, the *dc* magnetic cloak is theoretically investigated [12] and experimentally realized using superconductors and ferromagnetic materials [13,14]. By using the same materials as *dc* magnetic cloak, the theoretical realization of a *dc* magnetic concentrator is also demonstrated [15]. On the basis of form invariance of the heat conduction equation, transformation thermodynamics is investigated to manipulate diffusive heat flow [16]; through tailoring inhomogeneity and anisotropy of conductivities, transient thermal cloaking has been experimentally demonstrated [17]. In addition, manipulation of heat flux with only two kinds of materials (by utilizing a multilayered composite approach) has been reported [18,19]. Recently, a transformation-optics based *dc* electric cloak, composed of inhomogeneous and anisotropic conductivities, has been implemented using anisotropic and spatially-varying network of resistors [20]. Soon after, an ultrathin *dc* electric cloak [21] and a *dc* electric concentrator [22] are reported using similar resistor networks. More recently, an exterior *dc* cloak [23] and an active *dc* cloak [24] have been experimentally realized by the use of inhomogeneous and anisotropic conductivities with the aid of active sources. It is noted that EM invisibility cloaks made of L-C networks have been experimentally demonstrated [25].

Here, we demonstrate the designs of novel devices (viz., cloaking sensor, bilayer cloak and fan-shaped concentrator) for manipulating *dc* currents with natural bulk materials, and we further experimentally realize the bilayer cloak and fan-shaped concentrator to confirm the proposed methodology. The significance of this work is twofold. First, only the most common bulk materials are employed to construct the proposed devices; this does not involve exotic materials that need to be mimicked with complicated resistor networks [20-24], thus pushing the transformation devices a big step further towards practical applications. Second, derived rigorously from electric conduction equation, our schemes are exact rather than approximate ones. Furthermore, the proposed devices can be arbitrarily scaled up and down without changing the materials.

We begin with the concept of cloaking sensor in *dc* currents (which will naturally lead to the design of a bilayer cloak). Cloaking sensor [26] is a sensor wrapped by a shell that is capable of receiving incoming signal without distorting the external field. Fig. 1(a) demonstrates the concept of cloaking sensor in *dc* currents. We consider a round sensor (with radius of $b$) wrapped by a shell (with thickness of $c-b$). The conductivities of the sensor and shell are $\sigma_1$ and $\sigma_2$ respectively. A uniform *dc* current conducts along *x*-direction with current density of $J_0$, which is equivalent to a uniform external electric field $E_0$ applied in the *x*-direction due to $\vec{J} = \sigma \vec{E}$. Since the electric potentials satisfy Laplace's equation $\nabla^2 \phi = 0$ in all regions of space, they can be generally expressed as

$$\phi_i = \sum_{m=1}^{\infty} \left[ A_m^i r^m + B_m^i r^{-m} \right] \cos m\theta \tag{1}$$

where $A_m^i$ and $B_m^i$ ($i$=1, 2, 3) are constants to be determined by the boundary conditions and $\phi_i$ denotes the potential in different regions — $i = 1$ for the cloaking region (where $r \leq b$), $i = 2$ for the cloaking shell (where $b < r \leq c$) and $i = 3$ for the exterior region (where $r > c$).

Taking into account that $\phi_3$ should tend to $-E_0 r\cos\theta$ when $r\to\infty$, we only need to consider *m*=1. Since $\phi_1$ is limited when $r\to 0$, we can infer that $B_1^1 = 0$. As the electric potential and the normal component of electric field vector are continuous across the interfaces, we have

$$\begin{cases} \phi_i\big|_{r=b,c} = \phi_{i+1}\big|_{r=b,c} \\ \sigma_i \dfrac{\partial \phi_i}{\partial r}\bigg|_{r=b,c} = \sigma_{i+1}\dfrac{\partial \phi_{i+1}}{\partial r}\bigg|_{r=b,c} \end{cases} \quad (2)$$

Here, $\sigma_3 = \sigma_b$, where $\sigma_b$ is the electric conductivity of the background. By substituting Eq. (1) into Eq. (2), we obtain

$$B_1^3 = E_0 c^2 \frac{\sigma_2(Q_1-Q_2)-\sigma_b(Q_1+Q_2)}{\sigma_2(Q_1-Q_2)+\sigma_b(Q_1+Q_2)} \quad (3)$$

where $Q_1 = c^2\left(1+\dfrac{\sigma_1}{\sigma_2}\right)$ and $Q_2 = b^2\left(1-\dfrac{\sigma_1}{\sigma_2}\right)$. By setting $B_1^3 = 0$, we obtain

$$c = b\sqrt{\frac{(\sigma_2-\sigma_1)(\sigma_2+\sigma_b)}{(\sigma_2+\sigma_1)(\sigma_2-\sigma_b)}} \quad (4)$$

Obviously, cloaking sensor can be successfully achieved as long as Eq. (4) is fulfilled. Considering that the sensor (central region), cloaking shell, and background are stainless steel (with $\sigma_1 = 1.3\times 10^6$ S/m), copper (with $\sigma_2 = 5.9\times 10^7$ S/m), and iron (with $\sigma_3 = 1\times 10^7$ S/m), respectively, we obtain *c* = 2.9 cm when *b* = 2.5 cm according to Eq. (4). The simulated plot reproduced in Fig. 1(b) affirms the concept of cloaking sensor in *dc* currents compared to the case of bare sensor in Fig. 1(c).

The drawback of cloaking sensor is that the cloaking shell has to be changed when either the geometrical size or material of sensor is changed. Actually, in most cases, we only need to render an object invisible without receiving the incoming signal. Analogous to EM cloaks that prohibit incident waves with a PEC layer, an insulating layer (*a*<*r*<*b*) located between the object and cloaking shell may prevent the electric current from touching the object. We thus derive a bilayer cloak as conceptually demonstrated in Fig. 1(d). By setting $\sigma_1 = 0$, we obtain

$$c = b\sqrt{\frac{\sigma_2 + \sigma_b}{\sigma_2 - \sigma_b}} \qquad (5)$$

Hence, an exact bilayer cloak, derived directly from the electric conduction equation, has been obtained. Eq. (5) implies that the geometrical size of the bilayer cloak ($b$ and $c$) can be tuned at will without changing the materials of the outer layer and background (i.e., $\sigma_2$ and $\sigma_b$ are fixed).

The experimental realization of a bilayer cloak is schematically illustrated in Fig. 2 (a), which is composed of an insulating layer (where $a<r<b$) and a copper shell (where $b<r<c$). Here, air is wisely chosen as in the sulating layer, and can be very thin due to its good insulating property. The geometrical parameters are $a$=2.5 cm, $b$=2.6 cm, $c$=2.66 cm. We consider the case where the central region (cloaking region) is connected to the ground. The calculated potential distribution of bilayer cloak is plotted in Fig. 1(e), in which the *dc* currents (electric-field lines) are also presented. As expected, the *dc* currents bend conformally around the cloaking region and restore exactly outside the cloak without distortion, thus rendering the object invisible. When the bilayer cloak is removed, the simulation result of bare object is demonstrated in Fig. 1(f), in which severe distortions of potential distribution and *dc* current lines can be clearly observed.

In the experimental setup, a gradually changing structure was adopted to transform circular equipotential lines to planar ones in the observation area. Fig. 2(b) shows the simulated potential distribution of a homogeneous gradually changing structure: it is clear that planar equipotential lines have been successfully obtained in observation area. The inner layer, outer layer, and background of fabricated bilayer cloak are air, copper, and stainless steel, respectively. Its geometrical parameters are chosen the same as before: $a$=2.5 cm, $b$=2.6 cm, and $c$=2.66 cm. The simulation result demonstrated in Fig. 2(c) agrees very well with the pure background in Fig. 2(b). In the experimental setup, a *dc* power supply with 1 V magnitude is used as the source, and the voltage is measured using a FLUKE 45 Dual Display Multimeter.

The cloaking performance can be evaluated by measuring the potential distribution along the observation lines (shown as black dotted lines in the insets of Fig. 3) near the bilayer cloak. In the ideal case, the observation lines are straight equipotential lines. The simulated and measured results are shown in Fig. 3, in which (a) and (b) correspond to the normalized potential distribution at the left observation line (where *x* = –2.8 cm) and the right observation line (where *x* = 2.8 cm) respectively. As expected, without the cloak, the presence of object strongly distorts the original equipotential lines. When the central region is wrapped by the bilayer cloak, both forward or backward scattering are eliminated and the potential profiles restore exactly to the original equipotential lines (represented by straight lines). It is clear that the measurement agrees well with simulation, which validates our design scheme.

To further demonstrate that our proposed scheme is robust, we design and fabricate a bilayer cloak with background of iron, as provided in Supplementary. An exact cloak has to satisfy two conditions: **(1)** the external field should be repelled from the cloaked region, and **(2)** the external field outside the cloak should be undisturbed (as if nothing is there). Our bilayer cloak completely fulfills this ideal case based on the existence of an exact solution from the conduction equation.

Conceptually demonstrated in Fig. 1(g), a *dc* electric concentrator can enhance the electric field and current density in a given region without distorting the external field. The concentrator can be divided into three regions: focusing region (where $r<a$), shell region (where $a \leq r \leq b$), and external region (where $r>b$). Both the focusing region and external region have the same electric conductivity $\sigma_b$. We assume that the electric conductivity of the shell region is homogeneous but anisotropic with $\sigma_r = 2^n \sigma_b$ and $\sigma_\theta = 2^{-n} \sigma_b$, where $n>0$. Considering that a uniform electric field $E_0$ is externally applied in the *x*-direction, the potential for the three regions can be obtained:

$$\phi_1 = -E_0\left(\frac{b}{a}\right)^{1-l} r\cos\theta, \quad \phi_2 = -E_0\left(\frac{b}{r}\right)^{1-l} r\cos\theta, \quad \phi_3 = -E_0 r\cos\theta \qquad (6)$$

where $l = \sqrt{\sigma_\theta/\sigma_r} = 2^{-n}$. Clearly, we obtain $\phi_1 = \phi_2 = \phi_3$ if $n = 0$ which represents free space. If $n \to \infty$, we obtain $\phi_1/\phi_3 = b/a$, which means that 100% concentrating efficiency is achieved. Fig. 1(h) shows the simulated potential distribution with the same geometrical parameters as those in the concentrating experiment: $a$=1.2 cm and $b$=6 cm. Obviously, the *dc* current is focused into the central region without distortion. As a reference, the simulation result of a pure background is also demonstrated in Fig. 1(i).

On the basis of effective media theory (EMT), homogeneous but anisotropic material may be practically realized by alternatively stacking two natural materials in the azimuthal direction; the experimental realization of such a fan-shaped concentrator is shown in Fig. 4(a) where copper wedges are periodically placed with air gaps. To quantitatively examine the concentrating efficiency of our proposed concentrator, we calculate $CE = \left|\phi|_{x=a} - \phi|_{x=-a}\right|/\left|\phi|_{x=b} - \phi|_{x=-b}\right|$. Interestingly, we note from Fig. 4(b) that the concentrating efficiency of a fan-shaped concentrator is comparable with an ideal anisotropic concentrator with $n$=4.5; here the concentrating efficiency of the ideal anisotropic concentrator is defined as $CE = (b/a)^{-l}$. Clearly, the concentrating efficiency of the fan-shaped concentrator is still above 80% even when $b/a$=100. Furthermore, the fan-shaped concentrator (employing only copper) always keeps the external potential distribution undisturbed no matter what the ratio $b/a$ is, as shown in Figs. 4(c)-4(e). We observe that, outside the concentrator, the potential distribution is not distorted, which is the same as those in the homogeneous materials. According to $\vec{J} = \sigma\nabla\phi$, we obtain the current density of the central region as $\vec{J}_{center} = \vec{J}_{background} \cdot CE \cdot b/a$, where $\vec{J}_{background}$ denotes the current density of the background. It is calculated that the current density of the concentrating region will be enhanced by more than 80 times when $b/a$=100.

We fabricated the fan-shaped concentrator in Fig. 4(a) with $a$=1.2 cm and $b$=6 cm, including 36 copper wedges and 36 air wedges. Analogous to the bilayer cloak, a gradually changing structure is also adopted to transform circular equipotential lines to planar ones in the observation area. An exact concentrator has to satisfy two conditions: **(1)** the external field outside the concentrator should be undisturbed, and **(2)** the electric field or current density should be focused into a smaller region. To examine the first condition, we measured and simulated the normalized potentials along the observation lines at $x = -7$ cm and $x = 7$ cm, corresponding to the dotted black lines in Fig. 5(a), respectively. It is apparent that the potential profiles are the same as if there were nothing there. To examine the second condition, we measured and simulated the potential distribution along the observation line at $y = 0$ as shown in Fig. 5(b). Compared to the pure background with linear potential distribution, the concentrator makes the voltage change sharply in the central region, which unambiguously demonstrates the concentrating effect. The measurement results agree well with simulation results, which confirm that our fan-shaped concentrator completely fulfills these two conditions.

In summary, we have demonstrated the manipulation of *dc* currents with natural bulk materials, and experimentally confirm the methodology through bilayer cloak and fan-shaped concentrator. Our design schemes do not rely on transformation optics, and we can thus avoid the problems present in previous proposals (such as inhomogeneous and anisotropic parameters that need to mimicked via complicated resistor networks [20-24]). Also, the proposed schemes, derived directly from the electric conduction equation, are exact rather than approximate ones. Finally, excellent performance can be achieved by employing only natural bulk materials, thus indicating that our advanced scheme may be readily extended for various applications beyond *dc* control [7,13,14,18,27,28].

**Acknowledgements**

TH, HY and YL contributed equally to this paper. C.W.Q. acknowledges the Grant R-263-000-A23-232 by National University of Singapore. T. H. acknowledges the support by the National Science Foundation of China under Grant No. 11304253.

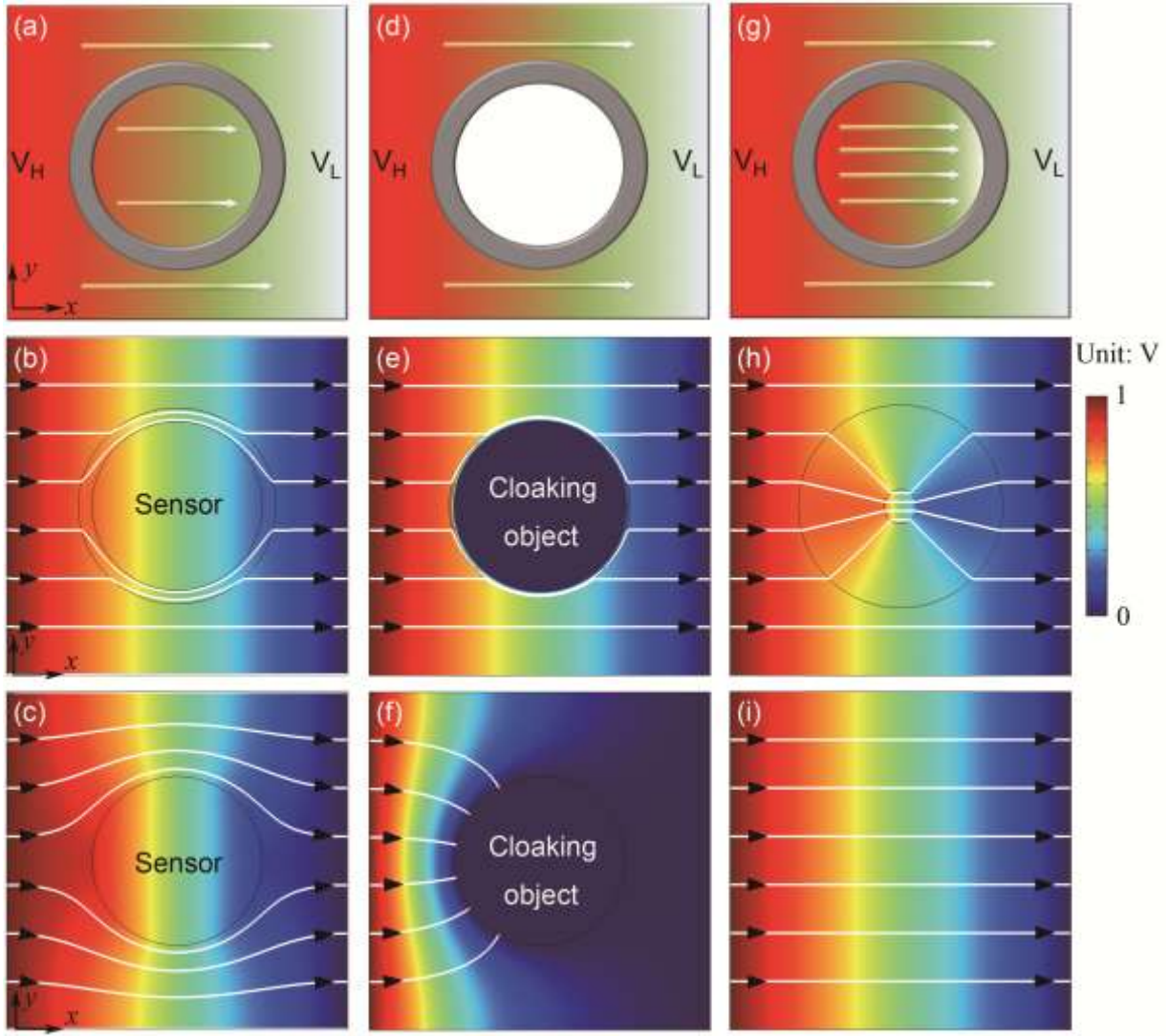

**Figure 1.** Demonstration of novel bilayer *dc* devices for manipulation of *dc* currents. (a) Cloaking sensor. (b) Simulation result of cloaking sensor. (c) Simulation result of a bare sensor. (d) *dc* bilayer cloak. (e) Simulation result of the bilayer cloak. (f) Simulation result of a bare and grounded object. (g) *dc* electric concentrator. (b) Simulation result of the concentrator. (c) Simulation result of a pure background without the concentrator. *dc* currents are represented with arrow lines in panel.

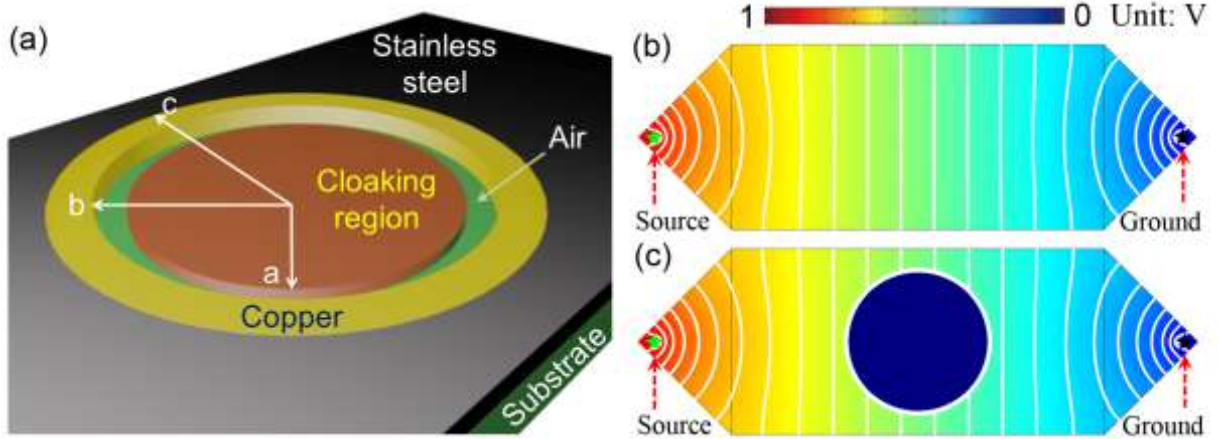

**Figure 2.** (a) Schematic illustration for experimental realization of *dc* bilayer cloak. (b) Potential distribution of a homogeneous gradually changing structure. (c) Potential distribution for the practical bilayer cloak in experiment. Equipotential lines are also demonstrated with white color in panel.

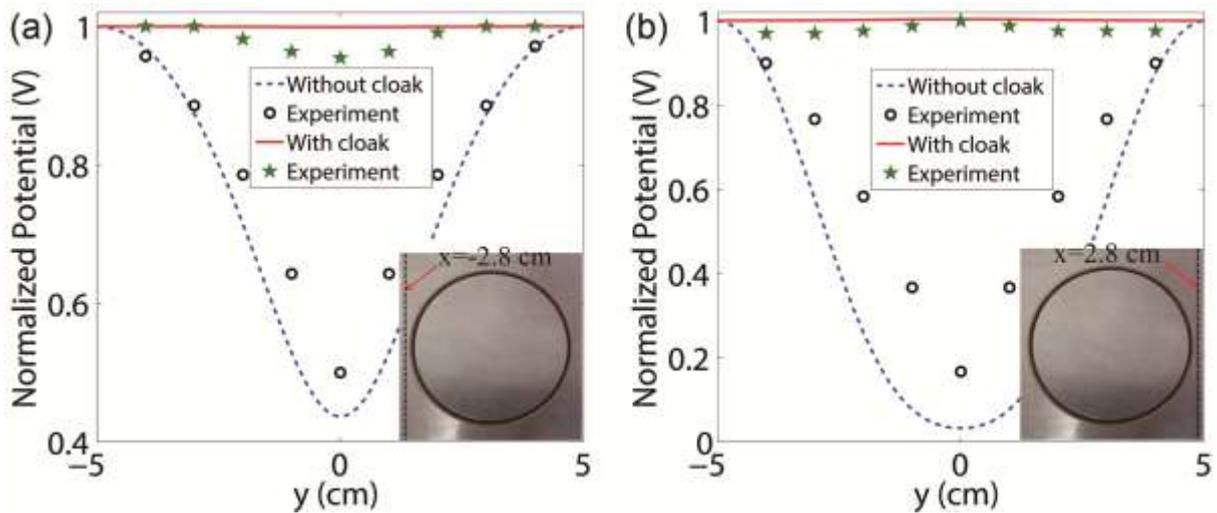

**Figure 3.** Simulation and experimental results of bilayer cloak with *a*=2.5 cm, *b*=2.6 cm, *c*=2.66 cm. (a) Normalized potential distribution at the left observation line *x*=-2.8 cm presenting backward scattering. (b) Normalized potential distribution at the right observation line *x*=2.8 cm presenting forward scattering. Insets denote the position of observation lines.

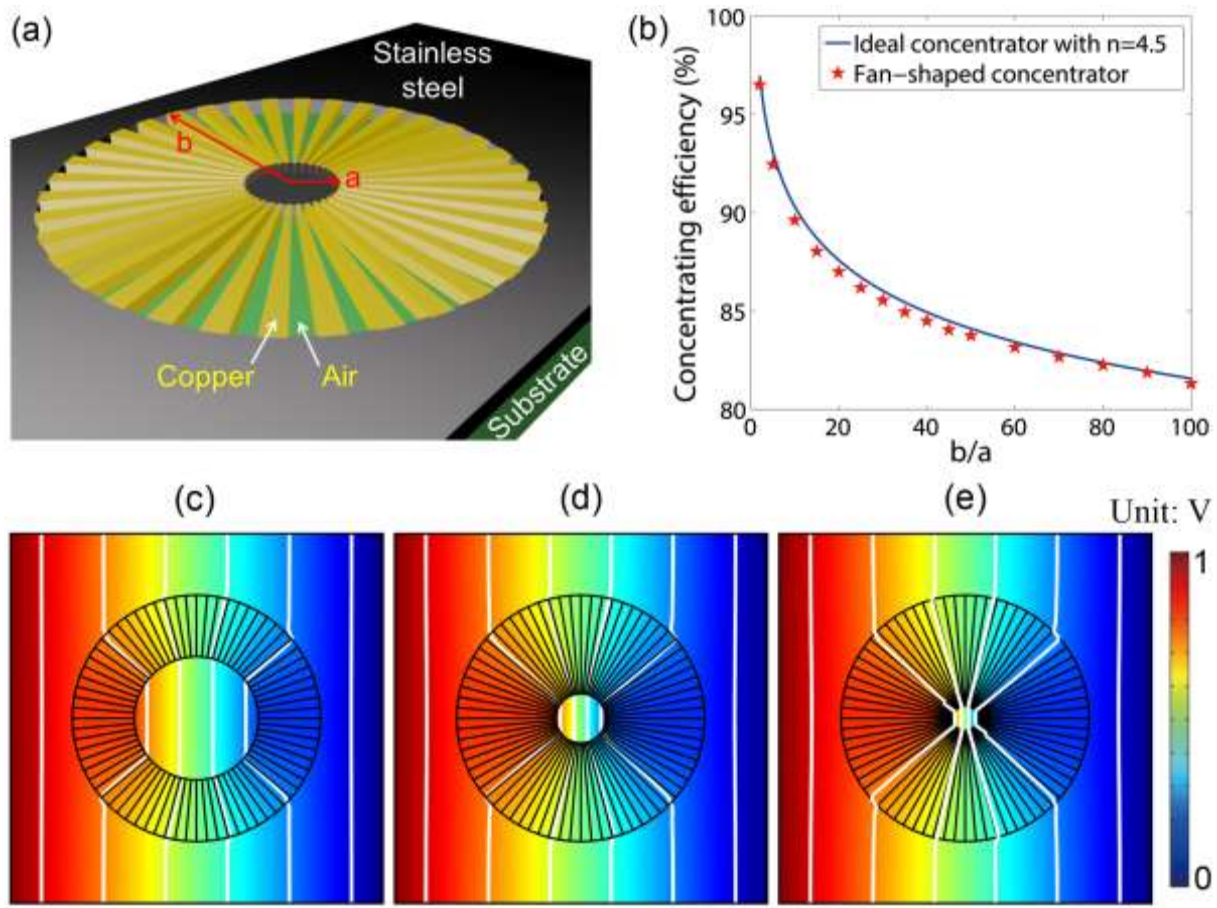

**Figure 4.** (a) Schematic illustration for experimental realization of fan-shaped concentrator. (b) Calculated concentrating efficiency for ideal concentrator with *n*=4.5 and fan-shaped concentrator in (a). Simulated potential distribution with different *b*/*a*: (c) *b*/*a*=2, (d) *b*/*a*=5, (e) *b*/*a*=10. Equipotential lines are also demonstrated with white color in panel.

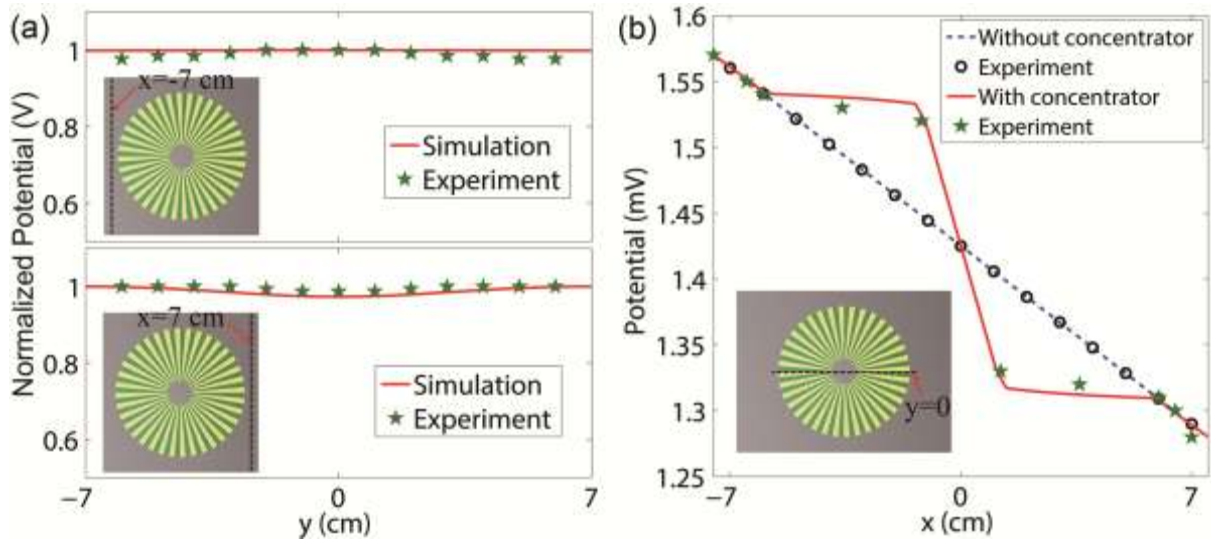

**Figure 5.** Simulation and measurement results of fan-shaped concentrator with $a$=1.2 cm and $b$=6 cm. (a) Normalized potential distribution at the observation lines $x$=-7 cm and $x$=7 cm, presenting backward scattering and forward scattering, respectively. (b) The simulated and measured potential distributions along a line $y$=0. Insets denote the position of observation lines.

# Manipulating *dc* currents with bilayer bulk natural materials

*Tiancheng Han, Huapeng Ye, Yu Luo, Swee Ping Yeo, Jinghua Teng, Shuang Zhang, and Cheng-Wei Qiu\**

\*       Dr. Tiancheng Han, Huapeng Ye, Prof. Swee Ping Yeo, Prof. Cheng-Wei Qiu

Department of Electrical and Computer Engineering, National University of Singapore, 119620, Republic of Singapore

E-mail: chengwei.qiu@nus.edu.sg

\*       Dr.  Yu Luo

Department of Physics, Blackett Laboratory, Imperial College London, London SW7 2AZ, UK

\*       Prof. Jinghua Teng

Institute of Materials Research and Engineering, Agency for Science, Technology and Research, 3 Research Link, Singapore 117602, Singapore

\*       Prof. Shuang Zhang

School of Physics and Astronomy, University of Birmingham, Birmingham B15 2TT, UK

Keywords: Cloaking sensor, *dc* bilayer cloak, fan-shaped concentrator, *dc* currents manipulation

**Bilayer cloak in Iron**

To demonstrate that our proposed scheme is robust, we design and fabricate a bilayer cloak with background of iron as shown in the photograph of Fig. S1. The inner layer and outer layer are still kept as air and copper. Its geometrical parameters are $a$=2.5 cm, $b$=2.6 cm, $c$=3.1 cm. Simulated results are presented in Fig. S2, in which (a) and (d) illustrate the cloaking region without and with the bilayer cloak, respectively. As expected, when the object is wrapped by the bilayer cloak, the equipotential lines and *dc* current lines outside the cloak restore exactly without distortion, thus rendering the object invisible. Figs. S2 (b) and S2 (c) show the simulated results of the object with only a single layer of air and copper, respectively; significant distortions can be clearly observed.

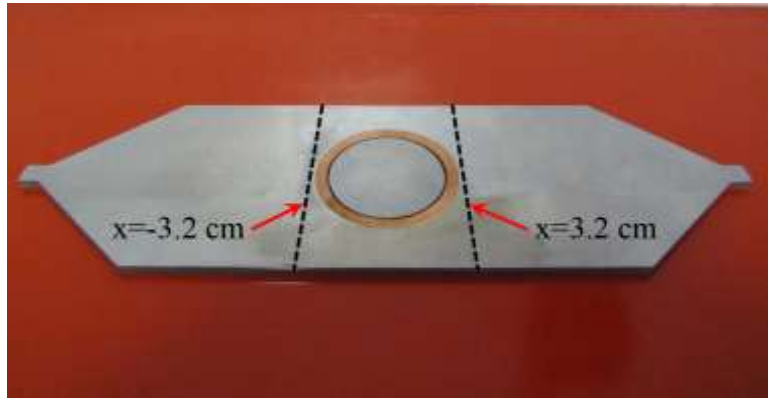

**Figure S1.** Photograph of fabricated bilayer cloak in Iron. Black dotted lines denote observation lines.

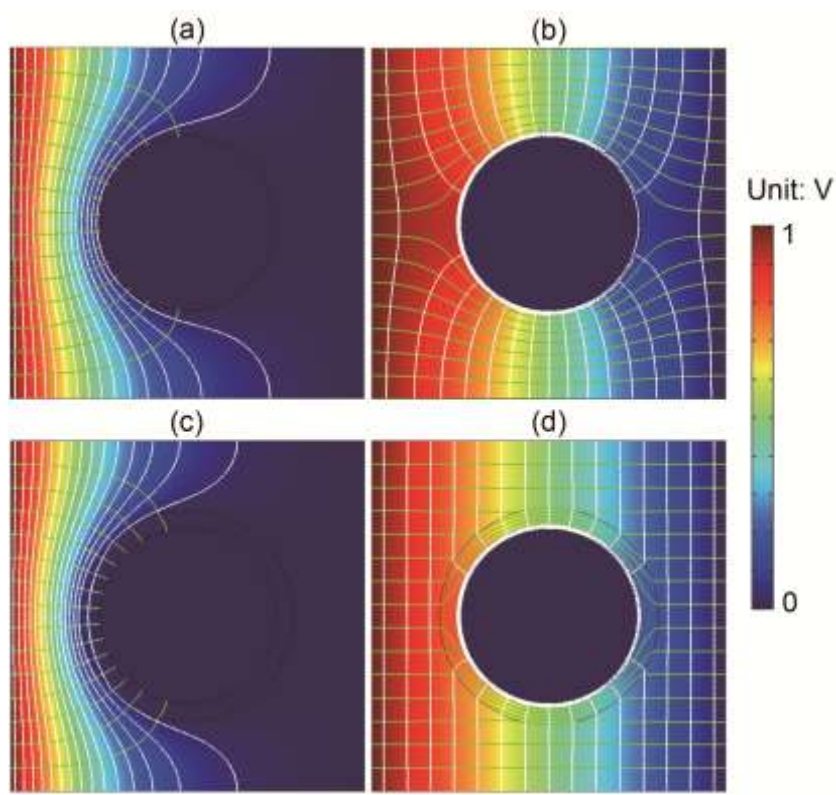

**Figure S2.** Simulation results for the bilayer cloak in iron with $a$=2.5 cm, $b$=2.6 cm, and $c$=3.1 cm. (a) A bare perturbation (object) exists. (b) The object is covered by a single air layer. (c) The object is covered by a single copper layer. (d) The object is wrapped by the proposed bilayer cloak. The central cloaking region (object) is connected to ground. Equipotential lines (white) and *dc* current lines (green) are also demonstrated in panel.

The measured results of bilayer cloak in iron are demonstrated in Fig. S3, in which (a) and (b) show the normalized potential distribution at the left observation line (where $x = -3.2$ cm) and the right observation line (where $x = 3.2$ cm) respectively. Again, the original equipotential lines are significantly distorted due to the bare object, and have been restored to original straight lines when the object is wrapped by our bilayer cloak. The measurement results agree very well with simulations. It is noted that the measured results of bilayer cloak in stainless steel (Fig. 3) is not as perfect as those in iron (Fig. S3), which is attributed to the fabrication errors because copper layer in stainless steel (0.06 cm) is much thinner than in iron (0.5 cm).

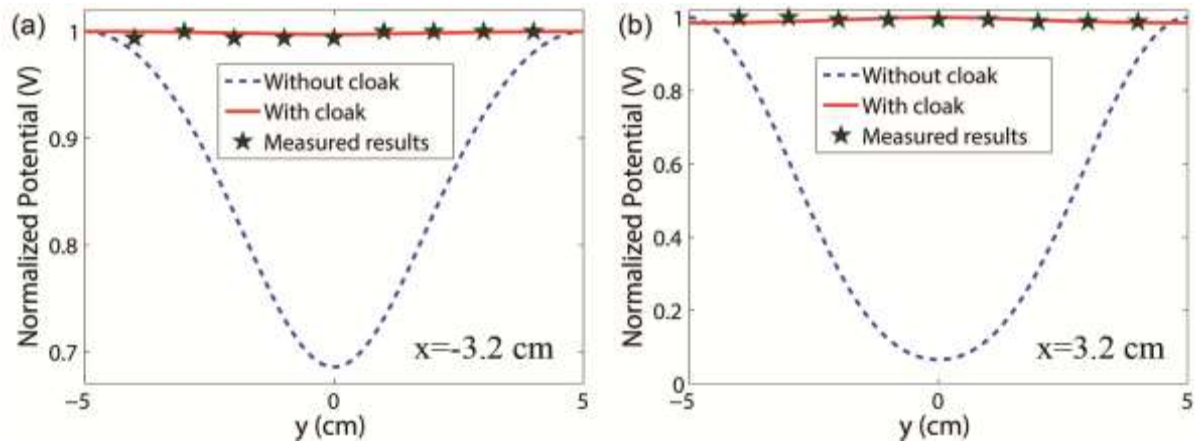

**Figure S3.** Simulation and experiment results of the bilayer cloak in Iron background with *a*=2.5 cm, *b*=2.6 cm, *c*=3.1 cm. (a) Normalized potential distribution at the left observation line *x*=-3.2 cm presenting backward scattering. (b) Normalized potential distribution at the right observation line *x*=3.2 cm presenting forward scattering.